\documentclass{anabs}
\usepackage{graphicx}
\usepackage{times}
\Volume{1-2}             
\Year{2003}              
\Month{02}               
\Pagespan{000}{000}      

\begin{document}
\lhead[\thepage]{A.N. Pradas, J., Kerp, J. \& Kalberla, P. M. W.: The soft X-ray background towards the northern sky. A detailed analysis of the Milky Way halo.}
\rhead[Astron. Nachr./AN~{\bf 324} (2003) 1/2]{\thepage}
\headnote{Astron. Nachr./AN {\bf 324} (2003) 1/2, 000--000}

\title{The soft X-ray background towards the northern sky. A detailed analysis of the Milky Way halo.}

\author{J. Pradas, J. Kerp \& P. M. W. Kalberla}
\institute{Radioastronomisches Institut der Universit\"at Bonn, Auf dem H\"ugel 71, 53121 Bonn, Germany}

\correspondence{jpradas@astro.uni-bonn.de}

\maketitle

\section{Radiative transport}
The correlation of the ROSAT all-sky survey (Snowden et al. 1995) with the Leiden/Dwingeloo \ion{H}{i} 21\,cm-line survey (Hartmann \& Burton 1997) leads to a consistent model of the soft X-ray background (SXRB). The diffuse X-ray radiation originates from three different components:  {\bf I.} The Local Hot Bubble (LHB) ($I_{\rm LHB}$), {\bf II.} The diffuse X-ray emission of the Milky Way halo 
($I_{\rm HALO}$), {\bf III.} The extragalactic X-ray background ($I_{\rm EXTRA}$). The tracer for the absorption is the \ion{H}{i} column density ($N_{\ion{H}{i}}$). The radiative transport equation is:

\begin{equation}
\label{eqn_radiation}
I = I_{\rm LHB} + I_{\rm HALO} \cdot {\rm e}^{- \sigma_\mathrm{h} \cdot N_{\ion{H}{i},{\mathrm h}}} + I_{\rm EXTRA} \cdot {\rm e}^{- \sigma_\mathrm{e} \cdot N_{\ion{H}{i},{\mathrm e}}}
\end{equation}

  In this work, we use this approach to model the SXRB across 50\% of the sky and the {\em whole} ROSAT energy window (0.1 to 2.4 keV). This allows us to study the spatial distribution of the diffuse X-ray emission of the Milky Way halo and the temperature of the Milky Way halo plasma accurately.

\section{Deriving model parameters}
We use the ROSAT R7 (1.05 to 2.04 keV) band to calibrate the extragalactic component of our model. The contribution of the LHB and Milky Way halo to the ROSAT R7 band is negligible. We assume that the intensity of the LHB is constant. The contribution of the LHB is determined towards the high \ion{H}{i} column density regions and within the low energy ROSAT bands. We tried to fit single and double (Kuntz \& Snowden 2000) halo plasma temperature models to the X-ray data in {\em all} ROSAT energy bands simultaneously. In order to model the {\bf variation of halo intensity with galactic coordinates} (Kappes, Pradas \& Kerp 2002) we use a Milky Way hydrostatic equilibrium halo model (Kalberla 2002).

\begin{figure}
{\includegraphics[scale=0.23]{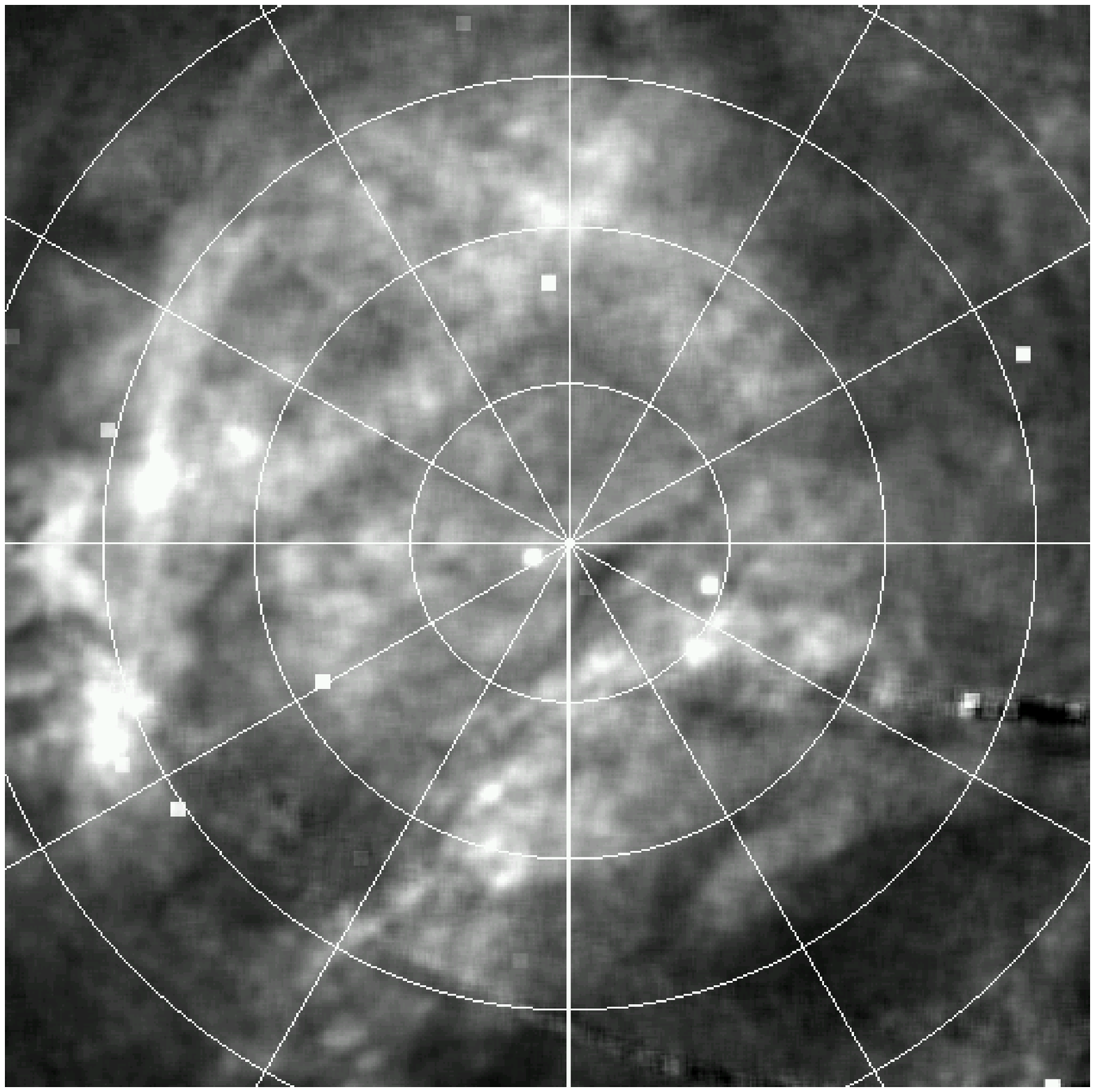}}
{\includegraphics[scale=0.23]{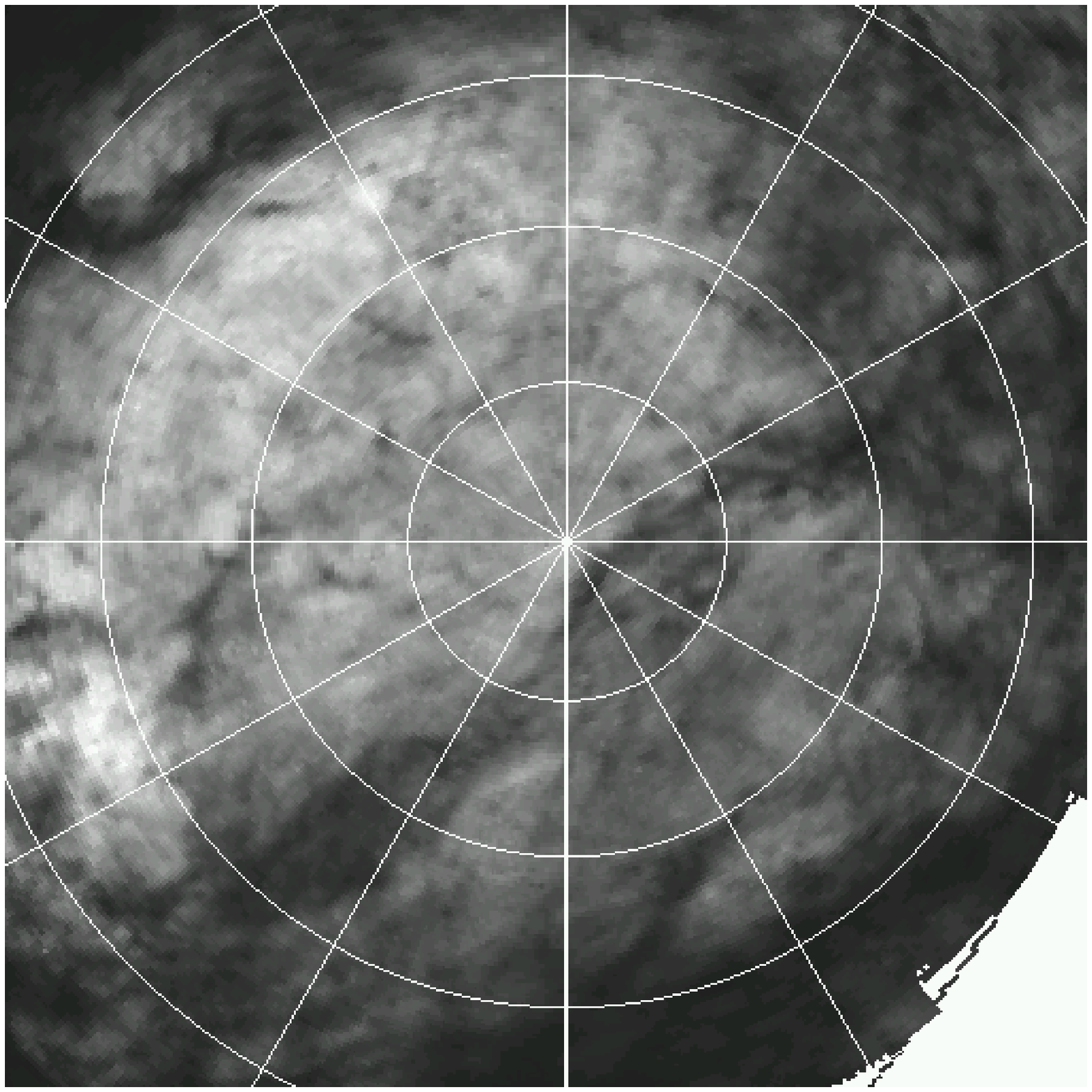}}
\caption{{\bf left:} The observed ROSAT R1 band intensity of a field centred at the north galactic pole. The step in galactic latitude is $15\degr$ and in galactic longitude is $30\degr$ with $l=0\degr$ directed to the bottom. {\bf right:} Modelled image of the same field. For an explanation for the regions showing discrepancies see Pradas \& Kerp (these proceedings)}
\label{xrhi}
\end{figure}

\section{Results}
After an iterative procedure to optimise $T_{\rm{HALO}}, T_{\rm{LHB}}, I_{\rm{HALO}}$ and $I_{\rm{LHB}}$, we conclude, that the galactic X-ray halo can be approximated with a {\bf single temperature plasma} with $10^{6.1}$ K $< T_{\rm{HALO}} < 10^{6.2}$ K whose intensity varies with galactic coordinates but not exceeding a factor of 3 across the northern sky. 

Moreover, individual coherent structures, where the model significantly deviates from the observations can be used as a {\bf distance indicator} to the absorbing matter (Pradas \& Kerp, these proceedings).

\begin{acknowledgements}
  The authors like to thank the Deutsches Zentrum f\"ur Luft- und
  Raumfahrt for financial support under grant No. 50 OR 0103.
\end{acknowledgements}

\end{document}